\documentclass[authoryear, showpacs, showkeys, 10pt, floatfix, nofootinbib]{article}

\usepackage[english]{babel}
\usepackage[utf8x]{inputenc}
\usepackage[T1]{fontenc}

\usepackage[a4paper,top=3cm,bottom=2cm,left=3cm,right=3cm,marginparwidth=1.75cm]{geometry}

\usepackage{graphicx}

\usepackage{epstopdf}
\usepackage{indentfirst}
\usepackage{amsmath}
\usepackage{amsfonts}
\usepackage{amssymb}

\usepackage{soul}

\usepackage[colorinlistoftodos]{todonotes}
\usepackage[colorlinks=true, allcolors=blue]{hyperref}

\usepackage{booktabs}
\def\x{\mathtt{x}}
\def\y{\mathtt{y}}

\def\zl{z_{\ell}\,}
\def\oml{\varpi_{\ell}}

\def\A#1{\mathcal{A}_{#1}}

\def\E#1{\mathcal{E}_{#1}}

\def\th#1{\theta_{#1}}

\def\ee#1{\times 10^{#1}}

\begin{document}

\title{Exact dynamical solution of the Kuramoto--Sakaguchi Model for finite 
networks of identical oscillators}

\author{Antonio Mihara and Rene O. Medrano-T \footnote{E-mails: {\tt mihara74@gmail.com, rene.medrano@unifesp.br}.} \\
Departamento de F\'{\i}sica, UNIFESP - Universidade Federal de S\~ao Paulo \\  
Rua S\~ao Nicolau, 210, 09913-030, Diadema  SP, Brazil}

\maketitle

\begin{abstract}

We study the Kuramoto--Sakaguchi (KS) model composed by any $N$ identical phase oscillators symmetrically coupled. Ranging from local (one-to-one, $R=1$) to global (all-to-all, $R=N/2$) couplings, we derive the general solution that describes the network dynamics next to an equilibrium. Therewith we build stability diagrams according to $N$ and $R$ bringing to the light a rich scenery of attractors, repellers, saddles, and non-hyperbolic equilibriums. Our result also uncovers the obscure repulsive regime of the KS model through bifurcation analysis. Moreover, we present numerical evolutions of the network showing the great accordance with our analytical one. The exact knowledge of the behavior close to equilibriums is a fundamental step to investigate phenomena about synchronization in networks. As an example, at the end we discuss the dynamics behind chimera states from the point of view of our results.

\end{abstract}

\maketitle

\section*{Introduction}

For more than forty years, the paradigmatic system of $N$ one-dimensional coupled phase oscillators, the Kuramoto model \cite{Kuramoto1975}, has been intensively studied to understand phenomena related to synchronization in biological, chemical, and electronic networks. Despite the simplicity of the dynamics of each oscillator ($\dot{\theta} = \omega$) strong efforts should be dedicated to find analytical solutions for a network of nonlinearly coupled oscillators, due to the high dimensionality of the system. Kuramoto showed a seminal solution giving rise to the prosper application of the mean--field theory in the Kuramoto model \cite{Kuramoto1984}. The method considers the network in the thermodynamic limit $N \to \infty$ with oscillators globally coupled. So that, the network is described oscillating with a mean frequency and its coherence is given by the magnitude of an order parameter. Since then this approach has been largely well succeeded in analytical investigations \cite{Kuramoto1984, Watanabe1993, Strogatz2000, Strogatz2001, Acebron2005, Ott2008, Omelchenko2014, Hu2014}.
In contrast, accurate results for the finite-size Kuramoto model remains a challenge due to the great number of equations involved, nevertheless, the dynamics is richer. While in the global coupling the full synchronization is the only stable equilibrium, in different topologies of the Kuramoto model multistability is allowed \cite{Wiley2006}. And, sustained by Lyapunov function argument,  the system would reach an equilibrium state as $t \to \infty$ \cite{Hemmen1993}. In this context, equilibriums play a central role in the network dynamics.

Multistability \cite{Tilles2011, Girnyk2012}, basin of attractions \cite{Delabays2017, Ha2012}, and traveling waves \cite{Hong2011} are some of fundamental phenomena directly related with equilibriums in variants of the Kuramoto model with both attractive and repulsive phase couplings \footnote{The dynamics of networks with repulsive phase couplings is almost unknown in spite its relevance in neurons networks. In the repulsive regime, the oscillators do not collapse in a single phase although they synchronize in frequency.}. These phenomena are  also observed in real-world networks \cite{Cohen1992, Ermentrout1994, Tsodyks1999, Newman2010}. Such manifestations are mostly studied in the continuous thermodynamic limit and keep not yet well understood. Exact solutions for lower number of oscillators in the Kuramoto model are mandatory in this study but they are still a topic of investigation \cite{Bronski2012, Wang2015}. 
In order to shed some light on those problems we study a class of the Kuramoto--Sakaguchi (KS) model \cite{Sakaguchi1986}, a generalization of the Kuramoto model explicitly for finite $N$. We obtain solutions that describe precisely the trajectories of each oscillators of the network when the system is close to an equilibrium. The collective  time evolution of the network can be followed by the order parameter \cite{Omelchenko2014} but this is the first time that the evolution of the network elements can be individually described analytically. We present several studies comparing numerical simulations with our theoretical predictions.

KS model is  given 
by
\begin{equation}
\dot{\theta}_{\x} = \omega_{\x} + \sum_{\langle\y,\x\rangle} G(\x, \y) 
\sin\left( \th\y - \th\x - \alpha \right) \, ,
\label{ks}
\end{equation}
where $\x = 0, 1, 2, ..., N-1$ identifies the $\x$-th oscillator in a ring, $\omega_{\x}$, its natural frequency, and $G(\x, \y)$, the coupling rule between it and the $\y$-th oscillator. The notation $\langle\y,\x\rangle$ means that the summation is on the $R$ nearest neighbors in both sides of oscillator $\x$, then $\y$ assumes the values $ \x-R, \x-R+1,...,\x, ..., \x+R$, where $R$ can be from $R=1$ ({\it local} coupling) to $R=(N-1)/2$ ({\it global} coupling), if $N$ is odd.   If $1 < R < (N-1)/2$  the coupling is called {\it nonlocal}. Finally, $\alpha\in(-\pi, \pi]$ is a constant that, for positive coupling, determines the regime of the network named {\it attractive} if $|\alpha|<\pi/2$ and {\it repulsive} otherwise.

Considering $N$ identical oscillators $(\omega_{\x} = \mbox{constant}, \forall\x)$ interacting according to KS model, with periodic boundary conditions, as in general studies of chimera states \cite{Abrams2004, Pikovsky2008, Motter2010} (for the case $\omega_{\x} \neq \omega_{\y}$ see \cite{Wang2015} ), an equilibrium is related to the phase difference between the nearest neighbors of oscillators $\Delta = \theta_{\x+1}-\theta_{\x}$. Due to the similarity, any homogeneous phase distribution of the oscillators along of a circle is an equilibrium, i.e., 
\begin{equation}
\Delta = \frac{2\pi}{N} q, \qquad q = 0, 1, 2, ..., N-1.
\label{eq1}
\end{equation}
The integer number $q$ denotes the number of loops needed to distribute the phase oscillators in the circle. The trivial solution is $\Delta = 0$ (or $2\pi$) corresponding to the full synchronization. For the rest ($q\neq 0$) we say that the network is synchronized in a {\it $q$-twisted state}. Note there are no different distribution for $q\geq N$. In this work we describe rigorously the stability properties of the $q$-twisted states assuming a general symmetric coupling $G(|\y - \x|) \equiv G_n$, only dependent on the absolute distance between oscillators, $n = \y-\x$. More specifically, we determine the set of eigenvalues associated to each state identifying the complete stability scenario of hyperbolic and non-hyperbolic equilibriums for a finite number $N$ of oscillators. As an application we show specifically this scenery in the parameter space $R \times q$ of the repulsive regime and calculate the bifurcation of twisted
states in the thermodynamic limit.
In the end we discuss how these results contribute to a dynamical interpretation of chimera states.

\section*{Results}

\subsection*{Synchronization frequency of $q$-twisted states}

Without loss of generality we assume that frequency $\omega_{\x} = 0$, periodic boundary conditions $(\th{0} = \th{N})$ and $N$ odd. Then Eq. (\ref{ks}) can be rewritten as 
\begin{equation}
\dot{\theta}_\x = \sum_{n =-R}^R G_n 
\sin\left( \theta_{\x+n} - \th\x - \alpha \right) \, .
\label{eq4}
\end{equation}

Based on the system symmetry, we assume that the network shall asymptotically converge to a $q$-twisted state with the same constant frequency $\dot{\theta}_{\x} =\Omega$. Assuming $\theta_0 = 0$, we obtain $\th\x = \Omega t + \Delta\x $.
Substituting this result in Eq. (\ref{eq4}), the network synchronization frequency $\Omega$ of any mode $q$ is obtained:
\begin{equation}
\Omega =  \sum_{n=-R}^R G_n \sin(n\Delta - \alpha) \, .
\label{eq44}
\end{equation}

\subsection*{Analysis of Stability}

We begin to analyze the stability of $q$-twisted states 
by taking into account a small perturbation in its solution :
$\th\x = \Omega t + \Delta \x + \E\x $. Then
\begin{equation}
\dot{\th\x} = \Omega + \dot{\E\x} \, 
\label{lhs}
\end{equation}

Substituting Eq.(\ref{lhs}) in the LHS of Eq.(\ref{eq4}) and expanding the
(perturbed) RHS of Eq.(\ref{eq4}) to first order in $\E\x$
we obtain (with $ \psi \equiv n\Delta - \alpha $)
\begin{equation}
\dot{\E\x} (t)  =  \sum_{n=-R}^R G_n 
\cos\psi \, ( \E{\x+n} - \E\x ) \, .
\label{eq:dotEx}
\end{equation}
With the ansatz $ \E\x (t) = \A\x e^{\lambda t}$, 
(or with vector notation: $\vec{\mathcal{E}} (t) = 
\vec{\mathcal{A}}\,  e^{\lambda t}$), 
Eq.(\ref{eq:dotEx}) becomes
\begin{eqnarray}
\!\!\!\!\!\!\lambda \A\x &=& \sum_{n=-R}^R \!G_n 
\cos(\psi)( \A{\x + n} - \A\x ) \nonumber\\
 &=& \sum_{n=1}^R G_n\left[ \rho_n \A{\x-n} + \mu_n \A{\x+n} \right]
 - \beta\A\x
\label{eigeneq}
\end{eqnarray}
where $\rho_n=\cos(n \Delta+\alpha)$, $\mu_n=\cos(n \Delta-\alpha)$,
and
\begin{equation}
 \beta = \sum_{n=1}^R G_n ( \rho_n + \mu_n ) =
 2\cos\alpha  \sum_{n=1}^R G_n \cos(n \Delta)
\end{equation}

The eigenvalue Eq. (\ref{eigeneq}) can be written in the matrix form 
$\lambda \vec{A} = M \vec{A}$, 
where $\vec{A} = [ \mathcal{A}_0, ..., \mathcal{A}_\x,..., \mathcal{A}_{N-1} ]^T$ and $M$ is a circulant matrix \cite{davis1979}
\begin{equation}
 M=
   \begin{pmatrix}
     -\beta & G_1 \mu_1 & G_2 \mu_2   & \cdots & G_2 \rho_2   & G_1 \rho_1 \\
     G_1 \rho_1  & -\beta & G_1 \mu_1 & \cdots &  G_3 \rho_3   & G_2 \rho_2 \\
     G_2 \rho_2  & G_1 \rho_1 & -\beta& \cdots &  G_4 \rho_4   & G_3 \rho_3 \\
         \vdots &   \vdots  & \vdots  & \vdots & \ddots & \vdots\\
     G_1  \mu_1 &  G_2 \mu_2  & G_3 \mu_3    & \cdots &  G_1 \rho_1 & -\beta
   \end{pmatrix}
 \end{equation}
 with (non-normalized) eigenvectors given by 
 \begin{eqnarray}
 \vec{A}_{\ell} &=& 
 \left[ \zl^0, \dots, \zl^{\x}, \dots, \zl^{N-1} \right]^T \, , \\
 &&\,\, \zl \equiv \exp\left( i\frac{2\pi}{N}\ell \right) \, , \nonumber
  \label{EiVectors}
 \end{eqnarray}
with $\ell = 0, 1, 2, ..., N-1$.

The $\ell$-th eigenvalue of the $q$--twisted
state is given by $\lambda_{\ell} = \gamma_{\ell} + i \oml$, with
\begin{equation}
\gamma_{\ell} = - 4\cos\alpha \sum_{k=1}^R G_k \cos\left(k\frac{2\pi}{N}q\right)
\sin^2\left(k \frac{\pi}{N}\ell\right)  
\label{eq:gamma}
\end{equation}
\begin{equation}
\oml = 2\sin\alpha \sum_{k=1}^R G_k \sin\left(k\frac{2\pi}{N}q\right)\sin\left(k \frac{2\pi}{N}\ell\right)\, .
\label{eq:omega}
\end{equation}

Any perturbation $\vec{\mathcal{E}} (t) $ can written in terms of 
the eigenmodes: $\vec{\mathcal{E}} (t) = \sum_{\ell} C_{\ell} \vec{F}_{\ell} (t)$,
with $\vec{F}_{\ell} (t) = \vec{A}_{\ell} e^{\lambda_{\ell}t}$ or,
without vector notation, $\E\x (t) = \sum_{\ell} C_{\ell}\, F_{\ell} (\x, t)$
where $F_{\ell} (\x, t) = \zl^{\x}\,e^{\lambda_{\ell}t}$ which is 
a wave function:
\begin{equation}
F_{\ell} (\x, t) =  e^{\gamma_{\ell}t}
\exp\left\{i\left[ \frac{2\pi\ell}{N}\x + \oml t  \right]\right\} \, .
\label{eq18}
\end{equation}

Notice that the eigenmode $\ell=0$, with eigenvalue $\lambda_0 = 0$ and eigenvector $\vec{A}_0 = 
 \left[ 1, 1, \dots, 1 \right]^T$, results
from the invariance of the system given Eq.(\ref{ks}) under a global phase shift: $\th\x
\rightarrow \th\x + C\, ,\,\, \forall \x$.
Then, despite of the $N$ equations following Eq. (\ref{ks}), the equilibriums are $(N-1)$-dimensional since they are related to the phase difference. In other words, the sum of all phase differences is a multiple of $2\pi$ due to the periodic boundary condition such that any phase difference can be obtained from the others $N-1$. 
Therefore the eigenmode $\ell = 0$ shall be disregarded below.

%#############################################################

\subsubsection*{Testing our results}

In order to test our results, we confronted them with some ``brute force'' 
numerical simulations. In one of many simulations we evolved a network of 
50 oscillators locally coupled ($R=1$) with $ \alpha = 1.5$ and $G_n=0.1$.
The time evolution of such a system was performed by direct numerical integration 
of the 50 differential equations (Eq.(\ref{eq4})) from random initial conditions
and the system reached a twisted state with $q=8$: the system
has frequency synchronization, i.e. all the oscillators
have the same phase velocity $\Omega = -0.20665$, and the phase difference
between any two neighbors is $\th\x - \th{\x-1} = 8 . (2\pi/50), \,\,\forall\x$. 

Regarding the theoretic aspect, for $q=8$ we can compute all
the (real part of) eigenvalues (only for $\ell>0$) with Eq.(\ref{eq:gamma}):
\begin{align*}
\gamma_1 = \gamma_{49} &\approx -0.598\ee{-4}, \\
\gamma_2 = \gamma_{48} &\approx -2.38\ee{-4}, \\
\gamma_3 = \gamma_{47} &\approx -5.32\ee{-4}, \\
& \cdots \\
\gamma_{24} = \gamma_{26} &\approx -151.\ee{-4}, \\
\gamma_{25} &\approx -152.\ee{-4} .
\end{align*}

We observe that all $\gamma_{\ell(>0)}$ are negative and, as 
we shall discuss in the next subsection, it is a clear signal that
the 8-twisted state is stable, in agreement with the numerical simulation. 
On the other hand if we assume that the lifetime 
of an eigenmode can be estimated as $\tau_{\ell} \sim |\gamma_{\ell}|^{-1}$,
clearly the eigenmodes $\ell = 1$ and 49 have the longest lifetimes.

For large times, but before the system reaches the final state $q=8$, it is 
reasonable to expect a behavior dominated by the eigenmodes $\ell=1,49$ of the 
8-twisted state: a wave (Eq. (\ref{eq18})) with amplitude decaying exponentially 
with constant $\gamma = \gamma_{1,49}$. Substituting the parameters of the network 
in Eqs.(\ref{eq44}, \ref{eq:omega}) we obtain, respectively, 
$\Omega_8 = -0.20665$ 
and $ \varpi_1 = | \varpi_{49} | = 0.0211 \equiv \omega_T $. 

\begin{figure}[hb]
\centering
   \includegraphics[width=0.99\linewidth]{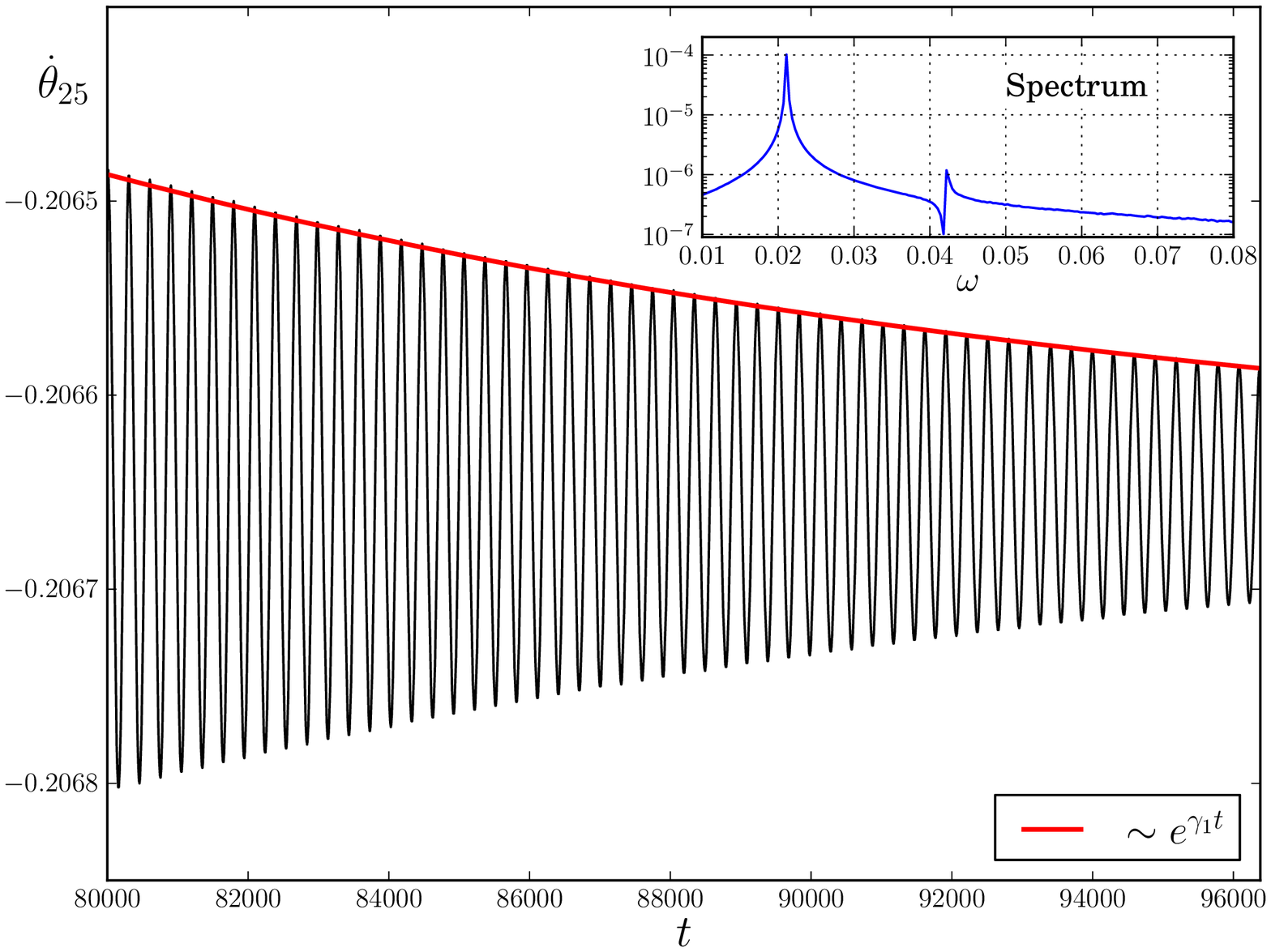} \\
\caption{Behavior of $\dot{\theta}_{25}$ in the KS model for 
local coupling (black). The exponential decreasing curve (red) and 
the spectrum (blue) demonstrate the behavior of the longest-lived eigenmodes
of the 8-twisted state. $N=50$, $R=1$, $\alpha = 1.5$, and $G_n = 0.1$.}
\label{figN50}
\end{figure}

In Fig.\ref{figN50} we show the behavior of $\dot{\theta}_{25}$ (the phase velocity 
of the oscillator in the site $\x = 25$) for $80000 \leq t \lesssim 96000$ (black line) 
and its corresponding spectrum.
One can observe in Fig.\ref{figN50} that the pronounced peak in the spectrum is around 
$\omega \approx \omega_T$ (blue line) and the 
oscillation amplitude of $\dot{\theta}_{25}$ is decaying exponentially
to the final synchronization frequency $\Omega_8$
as predicted above: the red curve is obtained with a function 
proportional to $\exp(\gamma t )$.

\subsection*{Some applications}

\subsubsection*{Stability of states in finite networks}

The eigenmodes\footnote{Remember that only the eigenmodes with $\ell>0$
are considered here.} are stable if $\gamma_{\ell}<0$ and unstable if $\gamma_{\ell}>0$. Thus, a $q$-twisted state can be classified as:
\begin{itemize}
\item[(a)] {\bf a}ttractor (hyperbolic) if  $\gamma_1, ..., \gamma_{N-1} <0$;
\item[(b)] {\bf r}epeller (hyperbolic) if  $\gamma_1, ..., \gamma_{N-1} >0$;
\item[(c)]  {\bf s}addle (hyperbolic) if $\exists\,\gamma_\ell \gamma_{\ell^{\prime}} <0$;
\item[(d)] {\bf n}on-hyperbolic if $\exists\,\gamma_\ell=0$.
\end{itemize}
We call the attention for some remarks about the 
items above. ({\it i}) If a non-hyperbolic state has all 
the other $\gamma_\ell < 0$ the system does not converge completely to the $q$-twisted state however it synchronizes. We call this state as neutrally stable. 
In the last subsection before the ``Discussion'' section we show with simulations
the different signatures of hyperbolic and non-hyperbolic stable twisted states: the former has a homogeneous phase distribution ($\th\x -
\th{\x-1} = $ constant, $\forall \x$) and the latter does not have.
({\it ii}) Due to the assumption of periodic boundary conditions
there is a symmetry between the stability of $-q$ and $q$ states: The real part of their eigenvalues [Eq. (\ref{eq:gamma})] are the same ($\gamma_q=\gamma_{-q}$). One also can see from 
Eq.(\ref{eq1}) that $-q$ and $N-q$ represent the same state,
so there is another way of labeling the twisted states, for 
instance:
$ q = -(N-1)/2,...,-1, 0, 1, ..., (N-1)/2$, if $N$ is odd.
({\it iii}) If $\alpha = \pi/2$ the eigenvalues are purely imaginary and the stability is not well defined by our assumptions. ({\it iv}) Since typically $\oml \neq 0$, there are several different types of equilibriums with stable and unstable manifolds which we generically call saddle. 
({\it v}) Although our deduction assumed $N$ odd, the analysis can be easily extended to $N$ even. 

\begin{figure}[h]
\centering
\includegraphics[width=0.7\linewidth]{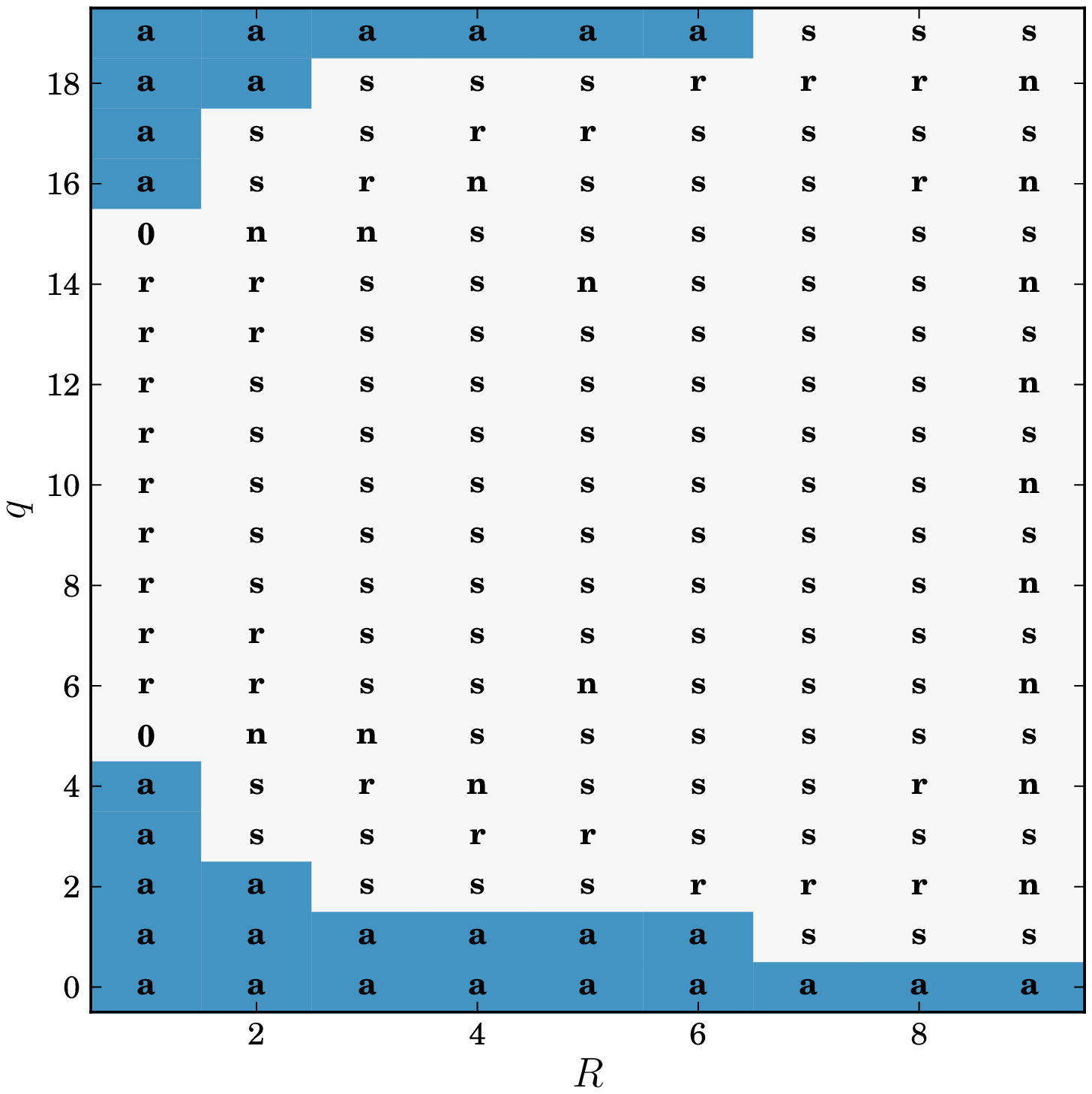} \\
\includegraphics[width=0.7\linewidth]{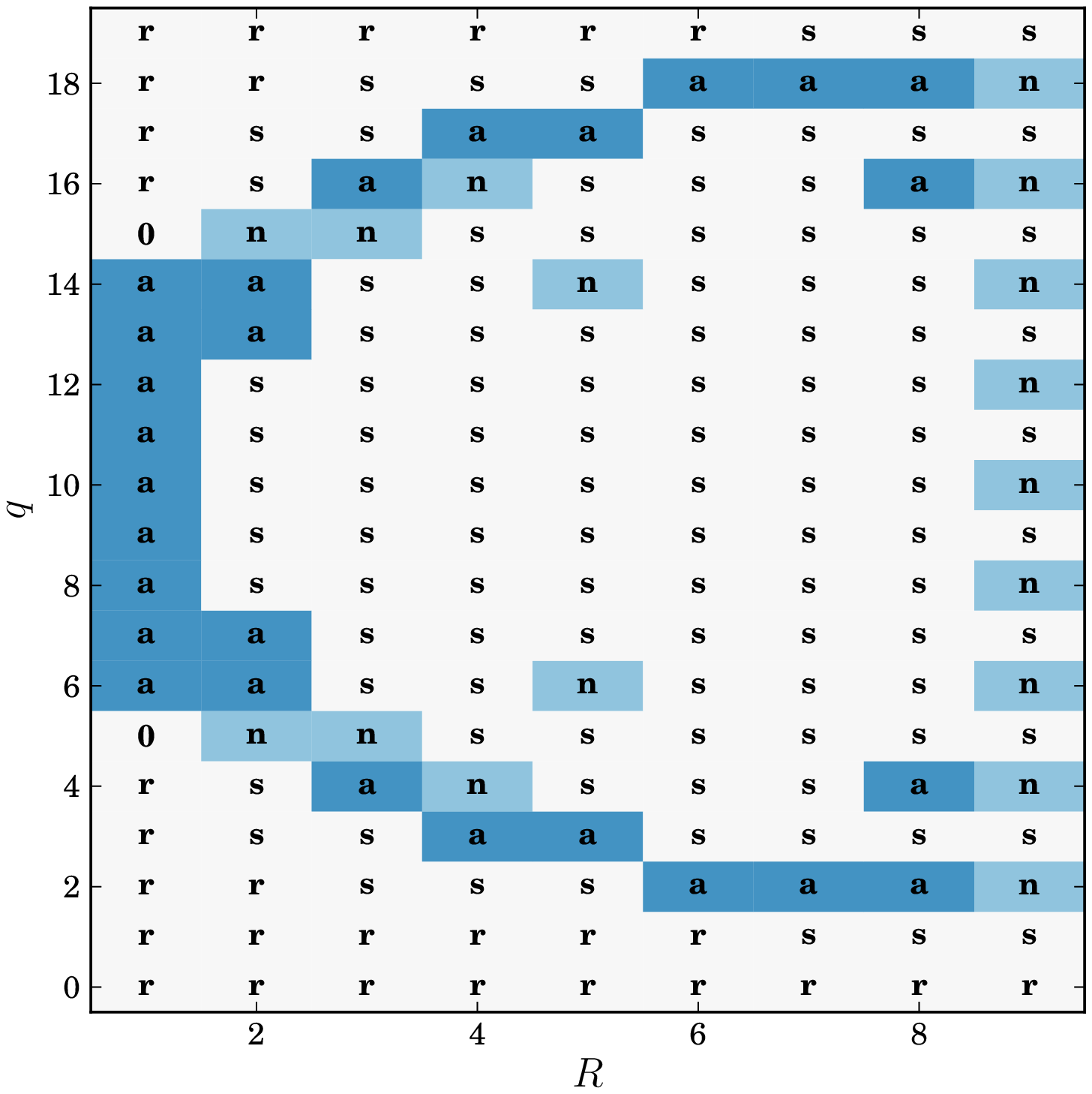}
\caption{Stability diagrams of the $q$-twisted states for $N=20$ oscillators equally coupled in a ring. Highlight for stable states: hyperbolic {\bf a}ttractors in blue, {\bf n}on-hyperbolic (or neutrally) stable states in light blue; {\bf s}addles, 
{\bf r}epellers and also {\bf n}on-hyperbolic unstable states are 
in white.
Top: ``attractive'' case ($\alpha=0$) and bottom: ``repulsive'' case ($\alpha=\pi$).}
\label{Stability20}
\end{figure}

As an example, we consider a network of $N=20$ oscillators in a ring with the same and positive coupling in the repulsive regime ($\pi/2<|\alpha|\leq \pi$). We varied $R$ from local to global and tested the stability of each $q$-twisted state according to the sign of $\gamma_{\ell}, \, \forall\, \ell > 0$. The outcome is compiled in Fig. \ref{Stability20}.

The stable states are highlighted in (light) blue in our diagrams and they 
could be compared with those in Fig. 1 of Ref.\cite{Girnyk2012}, which 
were obtained exclusively by direct numerical integration of the 
Kuramoto equations. The {\bf a}ttractors are represented by blue rectangles 
in the diagrams and the {\bf n}on-hyperbolic (or neutrally) stable states are 
in light blue.  

Due to Eq.(\ref{eq:gamma}) and conditions (a-d) above, if we change the 
system from ``attractive'' to ``repulsive'' regime (or vice versa) 
repellers become attractors and vice versa, including non-hyperbolic 
equilibriums, and saddles keep their unstable character in both regimes. 
The states with `0' have $\gamma_{\ell} = 0,\,\,\forall \ell$ and our 
analysis above is unable to determine their stability. 
However, according to our simulations, such states seem to be unstable in 
both regimes.

\subsubsection*{Bifurcations in the continuous limit}

Now we analyze the case of a network of oscillators in the continuum limit. 
We also assume below that each oscillator is coupled to its
$R$ nearest neighbors (each side) with constant coupling.

We start with a network locally coupled ($R=1$). From Eq. (\ref{eq:gamma}) the stability of a $q$-twisted state can change from attractor to repeller (vice versa) according to the sign of $\cos(2\pi q/N)$. For $|q|=N/4$ all eigenvalues are purely imaginary, consequently these states are not asymptotically stable or unstable. Therefore there is a bifurcation point given by the ratio  
$|q|/N = 1/4$.
 Since in our representation $|q|\leq (N-1)/2$, the maximum ratio $|q|/N$ tends to $1/2$ in the limit of $N \to \infty$. Therefore a network locally coupled 
presents approximately half of its states as (hyperbolic) attractors and 
the other half, repellers, independently of $\alpha$ and $G_n$.

For nonlocal couplings, the stability of the $q$-twisted states depends only on the  ratio $R/N$. This is clearly expressed in the thermodynamic limit $N\to\infty$ of Eq. (\ref{eq:gamma}) where, in the attractive regime, stable solutions are obtained when
\begin{equation} 
\int_0^{R/N} \cos(2\pi qz) \sin^2(\pi\ell z)\, \mathrm{d}z > 0
\label{eq20}
\end{equation}
 in the range $1 \leq R \leq N/2$, for $G_n > 0$ and constant. For a given  $q$-twisted state, its stability changes when the integral in the inequality $(\ref{eq20})$ is null. Defining the function $f(x,y)\equiv \sin(2\pi x y)/y$, with $x=R/N$, to express the solution of the integral, the bifurcation condition is given by
 \begin{equation}
 f(x,(\ell-q)) - 2f(x,q) + f(x,(\ell+q)) = 0 \, .
 \label{eq21}
 \end{equation}
For a chosen $q$, we set $\ell=1$ and vary $x$ from 0 to 0.5 and find values ($x_{q(\ell=1)}$) that solve the equation. This process should be repeated for $\ell = 2, 3, \dotsc$. Since the stability condition Eq.(\ref{eq20}) is satisfied when  $x\gtrsim 0$, a $q$-twisted state loses its stability when Eq. (\ref{eq21}) is satisfied for some $\ell$, remaining unstable in the rest of the range of $x$. In our example twist bifurcations happen for $\ell = 1$: $tb_{q(1)} = x_{q(1)}$ (See the middle column in table \ref{tab1}). This result is valid for more complex networks. In \cite{Wiley2006} the authors obtained the same solution from the mean-field approach considering a symmetric distribution of coupling $G_n$.
  
In the repulsive regime, the condition of stability is no longer satisfied when  
$x\gtrsim 0$ for non-local couplings. So now the first twist bifurcation point demarcates the boundary where a $q$-twisted state becomes stable. In this regime several twist bifurcation can occur as shown in the last column of table \ref{tab1}. For local couplings, condition becomes $|q|/N >1/4$.

\begin{table}[h]
    \let\mc\multicolumn
    \centering
    \begin{tabular}{ccc@{\hspace{0.2cm}}c@{\hspace{0.5cm}}c@{\hspace{0.2cm}}c}
        \toprule
          \mc2c{Twisted}     & \mc2c{Attractive} &  \mc2c{Repulsive} \\
          \mc2c{State}  & \mc2c{Regime {\bf (I)}} &  \mc2c{Regime {\bf (II)}} \\
          \cmidrule(r){1-2} \cmidrule(r){3-4} \cmidrule{5-6}
          &$q$  & $\ell$ & $tb_{q(\ell)}$ & $\ell$ & $tb_{q(\ell)}$ \\
        \midrule
         & 1   & 1   &  0.340461  & -    &  unstable         \\
         \hline
         & 2   & 1   &  1/6            & 5    &  0.277562         \\
          \hline
        & 3   & 1   &  0.110727  &   7  &  0.191433          \\
         &      &      &                   & 5    &  0.308065      \\
         \hline
         & 4   & 1   &  0.082948  & 9    &  0.145507          \\
         &      &      &                   & 7    &  0.225577         \\
         &      &      &                   &  2   &  5/12         \\
         \hline
          & 5   & 1   &  0.066323  & 11  &  0.117041       \\
          &      &      &                   & 9    &  0.178895          \\
           \bottomrule
    \end{tabular}  
    \caption{Bifurcations in the continuum limit. 
    }
    \label{tab1}      
\end{table}

\begin{figure}[hb]
\centering
    \includegraphics[width=0.9\linewidth]{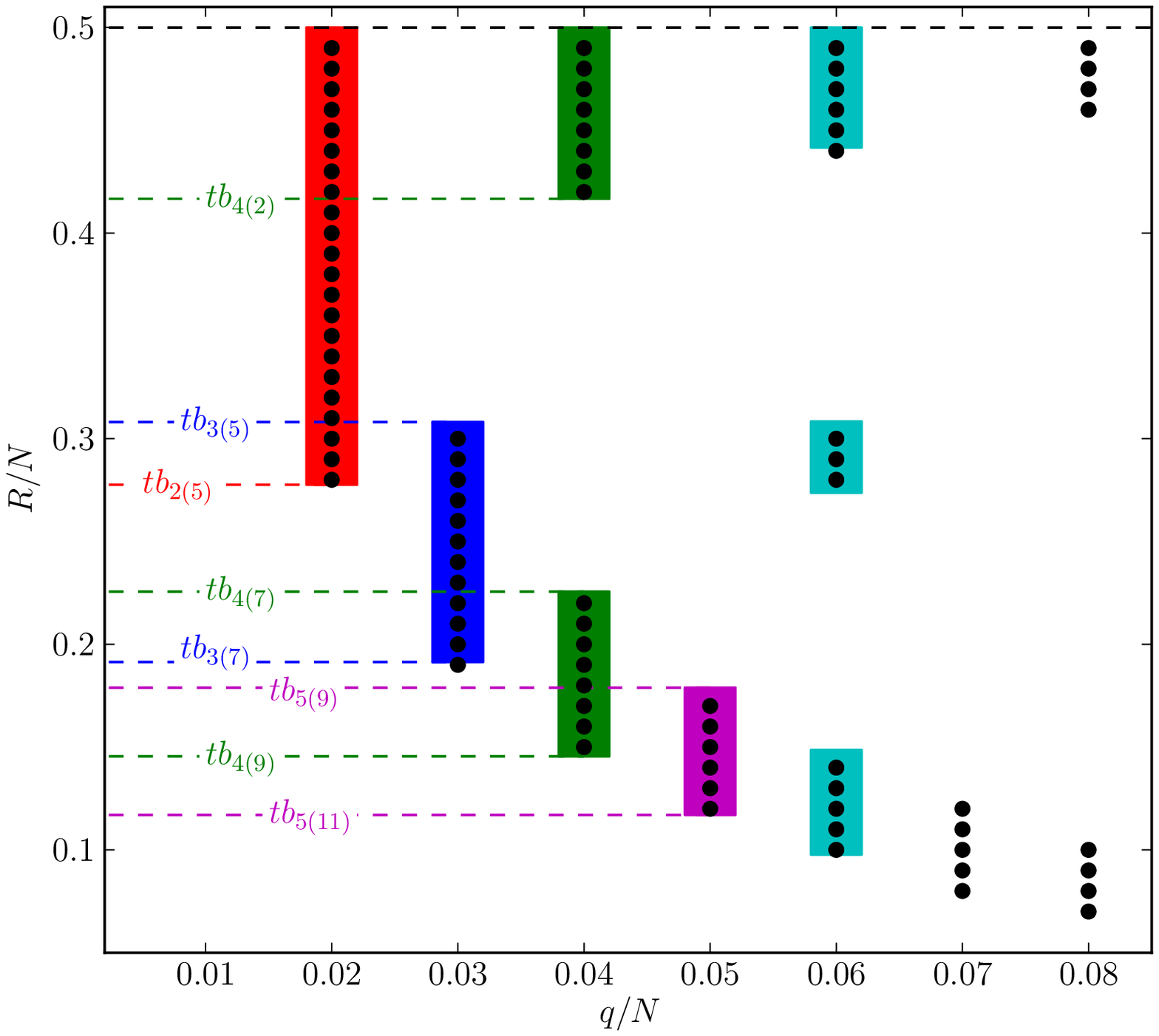}
\caption{Stability in the repulsive regime. Black dots are the 
stable states for $N=100$. Colored stripes are regions (in the continuum limit) 
where $q$-twisted states are expected to be stable, according to table \ref{tab1} 
(For comparison, the values of $q/N$ were calculated with $N=100$ in both cases.).  }
\label{fig2}
\end{figure}

For any value of $R/N$, the state of full synchronization ($q=0$)
shall be a hyperbolic attractor in the attractive regime
and a hyperbolic repeller in the repulsive regime. On the other 
hand the numbers in table \ref{tab1} should be interpreted 
as follows: {\bf (I)} In the attractive regime, $tb_{q(1)}$ is the highest value 
of $R/N$ where the $q$-twisted state is stable. For instance, $tb_{2(1)} = 1/6$ 
means that the state $q=2$ is stable for $0<R/N<1/6$. 
{\bf (II)} In the repulsive regime, several values of $tb_{q(\ell)}$ determine ranges of stabilities, for a given $q$. For instance: $q=4$ is stable when $tb_{4(9)}<R/N<tb_{4(7)}$ and $tb_{4(2)}<R/N<0.5$
(green stripes in Fig.\ref{fig2}) .

\subsubsection*{Time evolution of the Kuramoto order parameter}

It is important to remember that our analysis of stability were performed
by taking into account a small perturbation $\E{\x}$ in the $q$-twisted 
states. On the other hand with the (real part of the) eigenvalues 
Eq.(\ref{eq:gamma}) one can predict if a given $q$-twisted state is stable 
or not, then if the system is ``close'' to a stable state, 
the time evolution of each phase can be approximately
\begin{equation}
\th\x (t) \approx \Omega . t + \Delta . \x + \sum_{\ell} C_{\ell} F_{\ell} (\x, t) \, ,
\end{equation}
where $\Omega$ is the synchronization frequency of the state, given by Eq.(\ref{eq44});
$\Delta = 2\pi q/N$ and $F_{\ell} (\x, t)$ are the (perturbation) eigenfunctions
Eq.\ref{eq18}.
Eventually the time evolution of the Kuramoto order parameter, for a system close to an 
attractor, could be predicted directly with our formulas:
\begin{equation}
\rho(t) = \left|\frac{1}{N} \sum_{\x} e^{i\th\x (t)} \right| 
\label{eq:rho}
\end{equation}

We considered a system with the following parameters:
$\alpha=0.4,\, R=1,\,N=50,\, K=1$ and initially in the state $q_0=0$, 
(then $\Delta=0, \,\forall\,\x$), with a not too small
perturbation:
\begin{equation}
\theta_{\mathtt{x}} (0) = 1.4 e^{-(\mathtt{x}-25)^2/200} \sin( 7.5\pi \mathtt{x}/50 ) \, , 
\label{cond2}
\end{equation}
with $\x = 0, 1, ..., 49$.

\begin{figure}[ht]
\begin{center}
\includegraphics[width=1.0\linewidth]{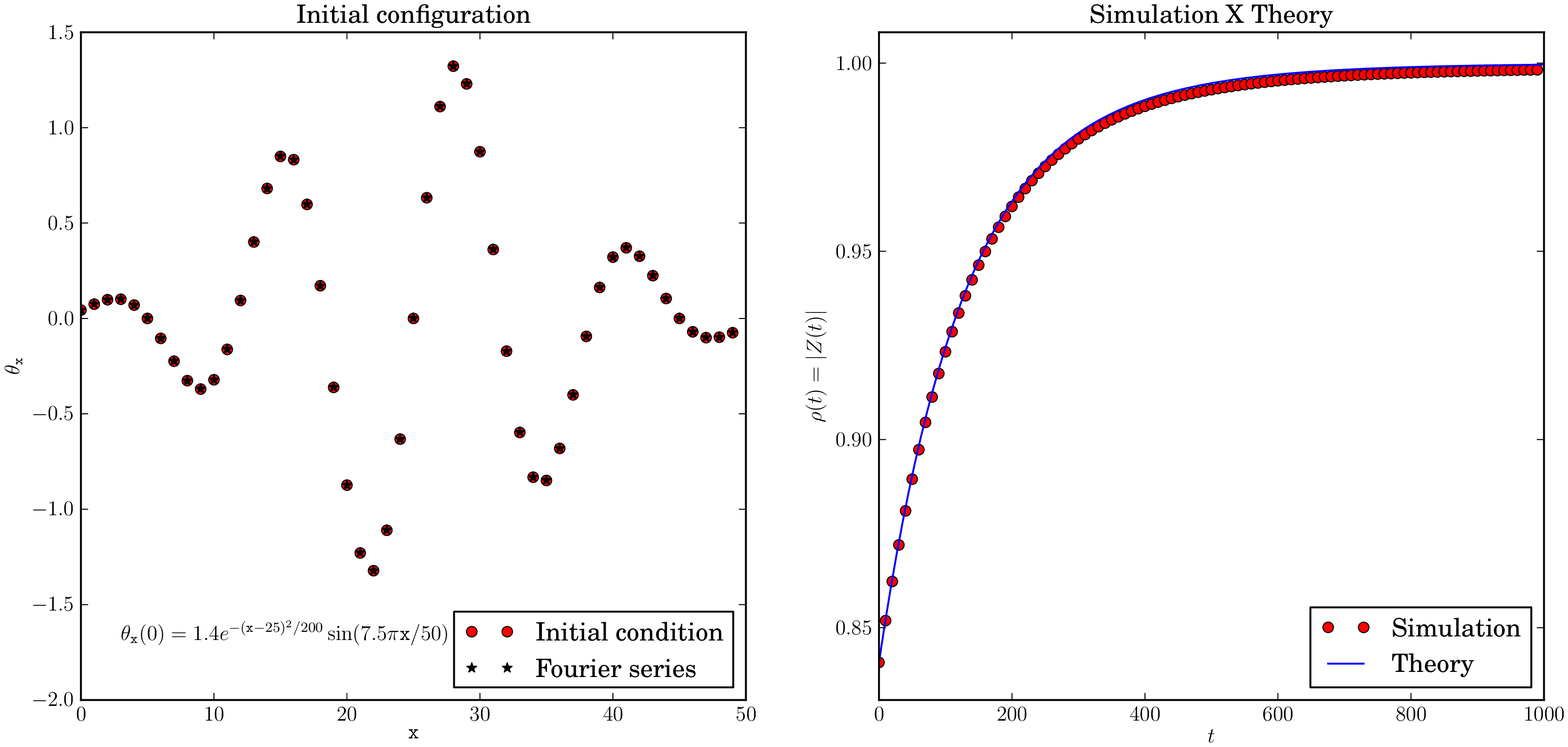}
\end{center}
\caption{Left: initial configuration.
Right: comparison between the theoretical
prediction (obtained with Fourier decomposition
and our equations for perturbations) and the results
obtained by ``Simulation'' = numerical 
integration of 50 differential equations. }
\label{fig:comparacao2}
\end{figure}

From the condition above we evaluated the phases by (i) ``brute force''
numerical integration of the 50 (KS) differential equations (``Simulation'') and
by (ii) our theoretical formulation: the initial condition Eq. (\ref{cond2}) 
was decomposed in its Fourier components $A_{\ell}, B_{\ell}$; $\Omega$ and 
the eigenvalues were obtained with Eqs. (\ref{eq44}, \ref{eq:gamma}, \ref{eq:omega})
for $q=0$; and the time evolution of each phase was calculated with:
\begin{eqnarray}
\th\x (t) =  \Omega t &+& \sum_{\ell} e^{\gamma_{\ell}.t} 
\left[ A_{\ell} \cos\left(
\frac{2\pi\ell}{50}.\x + \oml.t  \right) \right. \nonumber\\ 
&+& B_{\ell} \left. \sin\left( \frac{2\pi\ell}{50}.\x + \oml.t  
\right) \right] \, .
\end{eqnarray}
Then the Kuramoto order parameter was computed for both cases and
the comparison is presented in Fig. \ref{fig:comparacao2}.

\subsubsection*{Hyperbolic and Non-hyperbolic Stable States}

We show now the differences between hyperbolic and neutrally (or 
non-hyperbolic)
stable states with the results of some simulations. 

We simulated the case of a network with $N=20$ oscillators, $K>0$,
$\alpha = \pi$ (``repulsive'' regime), $R = 8$ and a initial
configuration ``close'' to $q = 2$:
\begin{equation}
\th\x (t=0) = \x . \frac{2\pi}{20} . 2 + r_{\x} \, , 
\quad \x = 0, 1, ..., 19 \, ,
\end{equation}
where $r_{\x}$ is a random number between -0.4 and 0.4.

By numerical integration (2nd order Runge--Kutta, with $\Delta t = 0.025$)
we measured the phase velocity ($d\th\x/dt$) of each oscillator, the phase
differences between oscillators and the mean value of $q$.

\begin{figure}[ht]
   \includegraphics[width=0.48\linewidth]{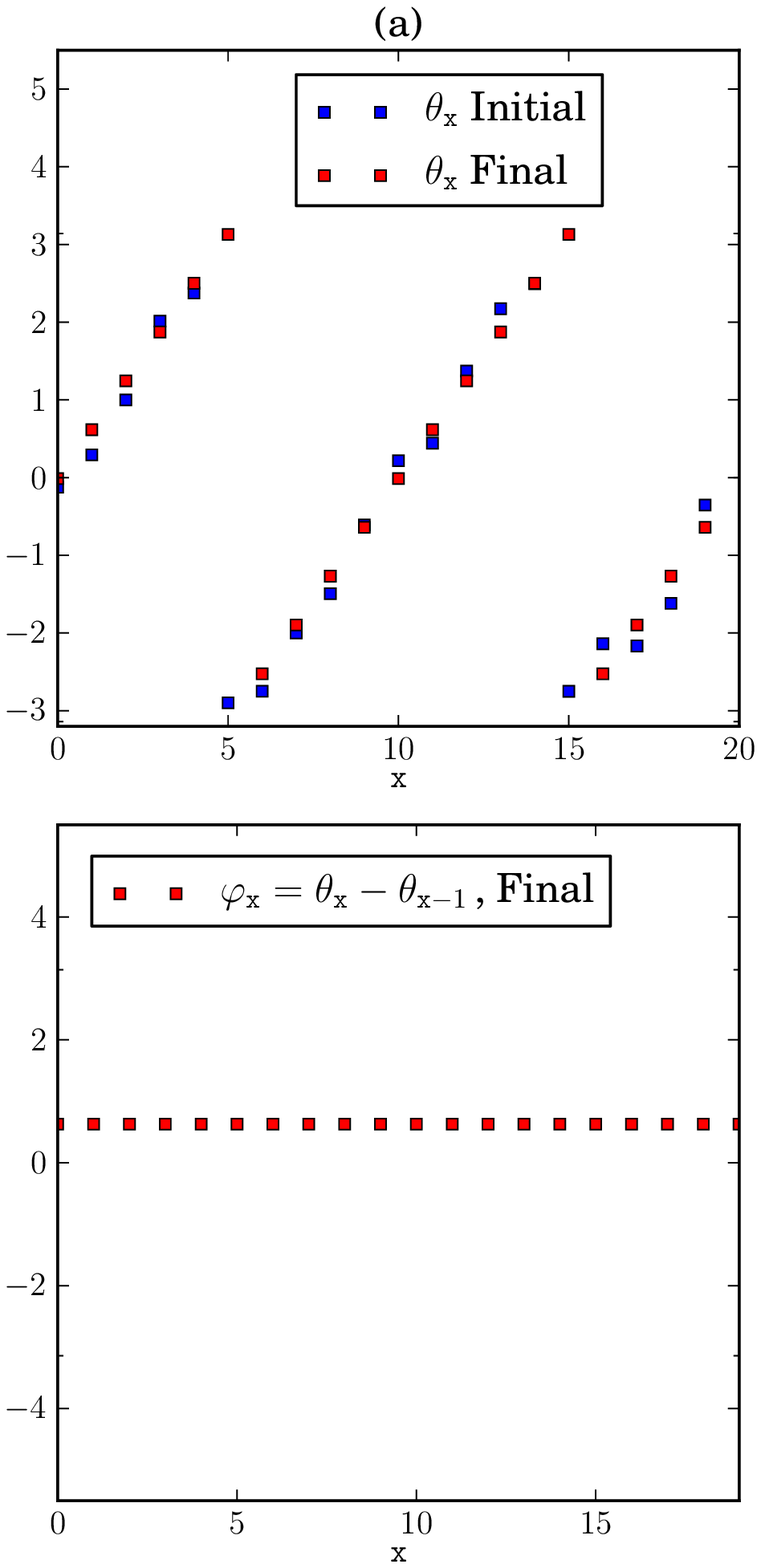}
   \includegraphics[width=0.48\linewidth]{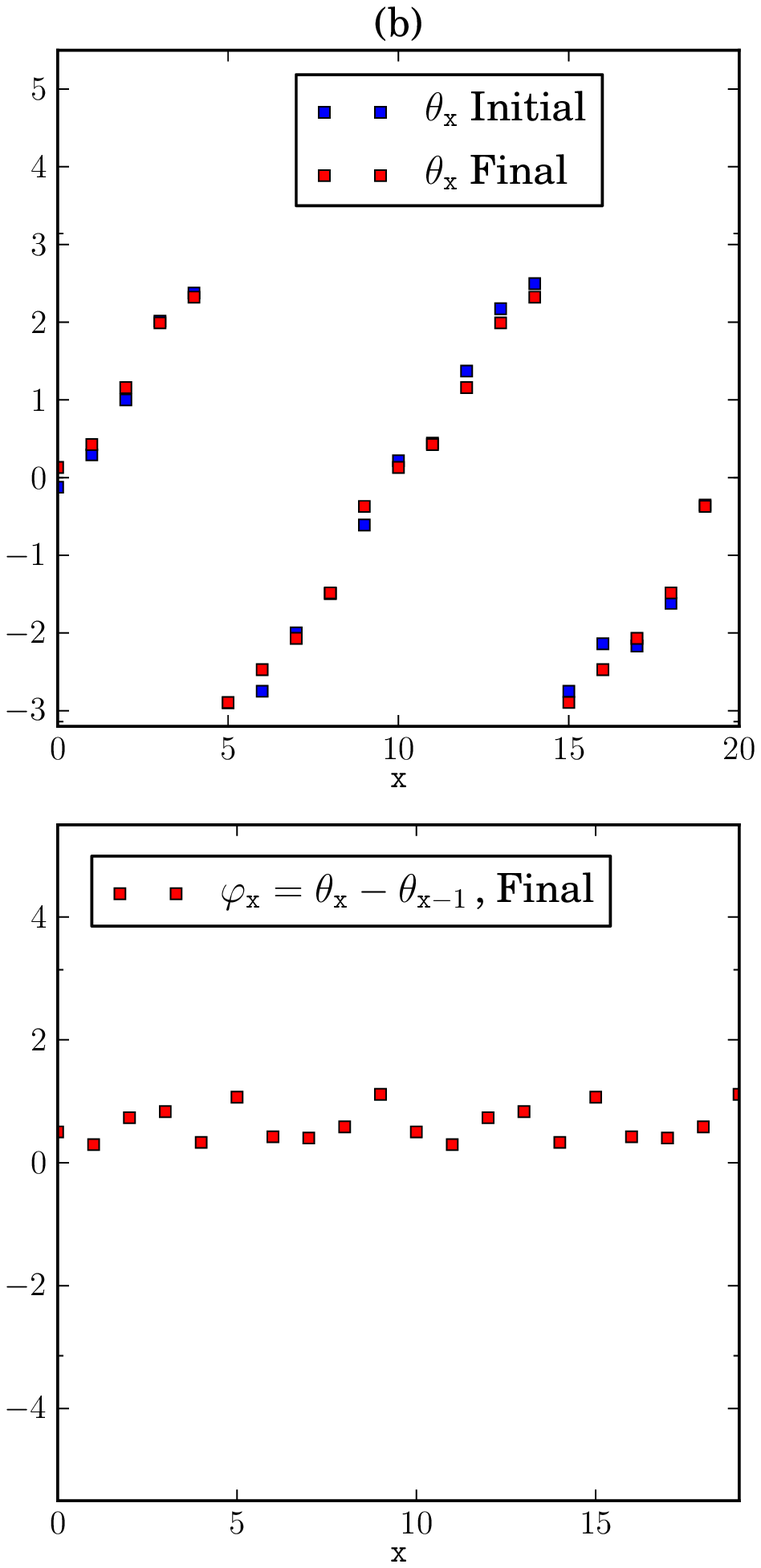}
\caption{ Signatures of the stable states in numerical simulations: 
(a) $(R=8, q=2)$ is a hyperbolic attractor: homogeneous distribution of phases;
(b) $(R=9, q=2)$ is a non-hyperbolic (or neutrally) stable state: non-homogeneous distribution of phases. }
\label{Estados}
\end{figure}

The results are presented in Fig.\ref{Estados}(a). We can observe in the
diagram of Fig.\ref{Stability20}(b) that $(R=8,q=2)$ is a {\bf hyperbolic} attractor. 
We observed that the system reaches a state in which
the distribution of phases is homogeneous: 
$(\th\x - \th{\x-1}) = \frac{2\pi}{N} q, \forall\x$, in this case
$N=20, q=2$. We can also observed that $\bar{q}=2$ during the entire simulation 
and each oscillator finished with null velocity.

On the other hand in the diagram of Fig.\ref{Stability20}(b) we can
notice that $(R=9,q=2)$ is a non--hyperbolic neutrally stable state. 
In order to observe the behavior of such a state we performed another 
simulation with almost the same conditions of the previous simulation, 
except with $R=9$, and the results are presented in 
Fig.\ref{Estados}(b). 

As occurred in the previous case, $\bar{q}=2$ during the entire simulation 
and each oscillator finished with null velocity. However the system 
does not reach a state with a homogeneous distribution of phases. 

%%%%%%%%%%%%%%%%%%%%%%%%%%%%%%%%%%%%%%%%%%%%%%%%%%%%%%

\section*{Final remarks}

In this work we presented the exact solution of the KS model for finite $N$ number of identical oscillators and give an extended analysis about the stability and the dynamics surround $q$-twisted states. So that, we are able to characterize precisely the nature around attractors, repellers, saddles and also non-hyperbolic equilibriums, some of the well known most fundamental invariant sets capable to dictate the global behavior in dynamic systems. We expect that this study gives new insights to understand several basic open problems in synchronization. Here we give an example. It is known that in KS models a chimera state (simultaneous manifestation of coherent and incoherent states in a network) collapses to $q=0$ (full synchronization) state after some time. In order to increase the lifetime without increasing the network, the relation $R/N \approx 0.35$ has been largely employed, seemingly empirically, in studies of chimera states in KS model and in several others different network models based on R{\"o}ssler, Lorenz, FitzHugh-Nagumo, Stuart-Landau, and Mackey-Glass dynamical systems \cite{Gopal2014, Zakharova2014, Rakshit2017}. 

According to our results, the bifurcations presented in table \ref{tab1} indicate that for $R/N \gtrsim 0.34$ only the state $q=0$ is stable. The other states are unstable, mostly saddles and some repellers. While these equilibriums contribute to the incoherent behavior, the equilibrium $q=0$ keep some oscillators synchronized. Additionally, if we enhance $N$ and $R$, but with fixed ratio $R/N \gtrsim 0.34$, there will appear new populations of saddles increasing the ergodicity 
of the environment and, consequently, extending the chimera lifetime. Such scenery seems to be in agreement with the previous numerical study in the KS model with $R/N\approx 0.35$ and $\alpha=1.46$ \cite{Wolfrum2011}. The authors 
also observed that the average lifetime of chimeras (before the full synchronization of the network) grows exponentially with $N$.

Another novelty from our results concerns to the repulsive regime of the KS model. Much attention has been paid to the full synchronization in networks, nevertheless bird flocks, fish schools, activity of cortical neurons in cats, and traveling waves in undulatory locomotion of fishes and lampreys are some of several manifestations in nature strongly related with a kind of synchronization where, contrary to the full synchronization and as in the repulsive regime of the KS model, the units  must not approach each other indefinitely. They converge asymptotically to different states resulting in a homogeneous distribution, crucial for the efficient operation of the network. It is impressive that despite of several important advances performed in the paradigmatic KS model its repulsive regime remained almost untouched. With our performed bifurcation study, this work arises the knowledge of the repulsive regime to the same level of the traditional attractive regime concerning the equilibrium states.

\section*{Acknowledgements}

R.O.M.T thanks M. Zaks and Y. Maistrenko for useful discussions, and acknowledge 
the support by S\~ao Paulo Research Foundation (FAPESP, Proc. 2015/50122-0). 
The plots were created with Python and its libraries: Matplotlib, Numpy and Scipy.

%\bibliography{sample}
%%%%%

\end{document}